
\overfullrule=0pt
\ifx\draft1\else\magnification 1200 \fi
\baselineskip=12pt

\hsize=16.5truecm \vsize=23 truecm \voffset=.4truecm
\parskip=14 pt
\def\idty{{\leavevmode{\rm 1\ifmmode\mkern -5.4mu\else\kern -.3em\fi I}}}
\def\Ibb #1{ {\rm I\ifmmode\mkern -3.6mu\else\kern -.2em\fi#1}}
\def\Ird{{\hbox{\kern2pt\vbox{\hrule height0pt depth.4pt width5.7pt
    \hbox{\kern-1pt\sevensy\char"36\kern2pt\char"36} \vskip-.2pt
    \hrule height.4pt depth0pt width6pt}}}}
\def\Irs{{\hbox{\kern2pt\vbox{\hrule height0pt depth.34pt width5pt
       \hbox{\kern-1pt\fivesy\char"36\kern1.6pt\char"36} \vskip -.1pt
       \hrule height .34 pt depth 0pt width 5.1 pt}}}}

\def\ibb #1{\leavevmode\hbox{\kern.3em\vrule
     height 1.5ex depth -.1ex width .2pt\kern-.3em\rm#1}}
\def\Nl{{\Ibb N}} \def\Cx {{\ibb C}} \def\Rl {{\Ibb R}}

\def\lessblank{\parskip=5pt \abovedisplayskip=2pt
          \belowdisplayskip=2pt }
\outer\def\iproclaim #1. {\vskip0pt plus50pt \par\noindent
     {\bf #1.\ }\begingroup \interlinepenalty=250\lessblank\sl}
\def\eproclaim{\par\endgroup\vskip0pt plus10pt\noindent}
\def\proof#1{\par\noindent {\bf Proof #1}\          
         \begingroup\lessblank\parindent=0pt}
\def\QED {\hfill\endgroup\break
     \line{\hfill{\vrule height 1.8ex width 1.8ex }\quad}
      \vskip 0pt plus2pt}

\def\Bar{\overline}
\def\abs #1{{\left\vert#1\right\vert}}
\def\bra #1>{\langle #1\rangle}

\def\dim {\mathop{\rm dim}\nolimits}

\def\ket #1{\vert#1\rangle}
\def\ketbra #1#2{{\vert#1\rangle\langle#2\vert}}

\def\set #1{\left\lbrace#1\right\rbrace}

\def\tr {\mathop{\rm tr}\nolimits}

\def\phi{\varphi}
\def\epsilon{\varepsilon}
\def\3{\ss}

\def\SU(#1){{\rm SU}(#1)}
\def\H{{\cal H}}
\def\maxent{{\cal M}}
\def\Flip{\Ibb F}
\def\ket#1>{\vert#1\rangle}
\let\up\uparrow  \let\down\downarrow

\def\phase{\chi}

\catcode`@=11
\def\ifundefined#1{\expandafter\ifx\csname
                        \expandafter\eat\string#1\endcsname\relax}
\def\atdef#1{\expandafter\def\csname #1\endcsname}
\def\atedef#1{\expandafter\edef\csname #1\endcsname}
\def\atname#1{\csname #1\endcsname}
\def\ifempty#1{\ifx\@mp#1\@mp}
\def\ifatundef#1#2#3{\expandafter\ifx\csname#1\endcsname\relax
                                  #2\else#3\fi}
\def\eat#1{}

\newcount\refno \refno=1
\def\labref #1 #2 #3\par{\atdef{R@#2}{#1}}
\def\lstref #1 #2 #3\par{\atedef{R@#2}{\number\refno}
                              \advance\refno by1}
\def\txtref #1 #2 #3\par{\atdef{R@#2}{\number\refno
      \global\atedef{R@#2}{\number\refno}\global\advance\refno by1}}
\def\doref  #1 #2 #3\par{{\refno=0
     \vbox {\everyref \item {\reflistitem{\atname{R@#2}}}
            {\d@more#3\more\@ut\par}\par}}\vskip\refskip }
\def\d@more #1\more#2\par
   {{#1\more}\ifx#2\@ut\else\d@more#2\par\fi}
\let\more\relax
\let\everyref\relax  
\newdimen\refskip  \refskip=\parskip
\let\REF\txtref
\def\@cite #1,#2\@ver
   {\eachcite{#1}\ifx#2\@ut\else,\@cite#2\@ver\fi}
\def\citeform#1{$^{#1}$}
\def\cite#1{\citeform{\@cite#1,\@ut\@ver}}
\def\eachcite#1{\ifatundef{R@#1}{{\tt#1??}}{\atname{R@#1}}}
\def\defonereftag#1=#2,{\atdef{R@#1}{#2}}
\def\defreftags#1, {\ifx\relax#1\relax \let\next\relax \else
           \expandafter\defonereftag#1,\let\next\defreftags\fi\next }

\newdimen\refskip  \refskip=\parskip
\def\@utfirst #1,#2\@ver
   {\author#1,\ifx#2\@ut\afteraut\else\@utsecond#2\@ver\fi}
\def\@utsecond #1,#2\@ver
   {\ifx#2\@ut\andone\author#1,\afterauts\else
      ,\author#1,\@utmore#2\@ver\fi}
\def\@utmore #1,#2\@ver
   {\ifx#2\@ut\and\author#1,\afterauts\else
      ,\author#1,\@utmore#2\@ver\fi}
\def\authors#1{\@utfirst#1,\@ut\@ver}

\let\everyref\relax            
\let\more\relax                
\let\reflistitem\citeform
\catcode`@=12
\def\Bref#1 "#2" #3\more{\authors{#1},\ {\it #2}\hfill\break (#3)\more}
\def\Gref#1 "#2"#3\more{\authors{#1}\ifempty{#2}\else:``#2''\fi,
                             #3\more}
\def\Jref#1 "#2"#3\more{\authors{#1}, #2, \Jn#3\more}
\def\JKref#1 "" #2\more{\authors{#1}, \Jn#2\more}
\def\inPr#1 "#2"#3\more{in: \authors{\eds#1}:``{\it #2}'', #3\more}
\def\Jn #1 @#2(#3)#4\more{{\it#1}\ {\bf#2} (#3)#4\more}
\def\author#1. #2,{#1.~#2}
\def\sameauthor#1{\leavevmode$\underline{\hbox to 25pt{}}$}
\def\and{, and}   \def\andone{ and}
\def\noinitial#1{\ignorespaces}
\let\afteraut\relax
\let\afterauts\relax
\def\etal{\def\afteraut{, et.al.}\let\afterauts\afteraut
           \let\and,}
\def\eds{\def\afteraut{(ed.)}\def\afterauts{(eds.)}}
\catcode`@=11

\newcount\eqNo \eqNo=0
\def\lasteq{\secNo.\number\eqNo}
\def\deq#1(#2){{\ifempty{#1}\global\advance\eqNo by1
       \edef\n@@{\lasteq}\else\edef\n@@{#1}\fi
       \ifempty{#2}\else\global\atedef{E@#2}{\n@@}\fi\n@@}}
\def\eq#1(#2){\edef\n@@{#1}\ifempty{#2}\else
       \ifatundef{E@#2}{\global\atedef{E@#2}{#1}}%
                       {\edef\n@@{\atname{E@#2}}}\fi
       {\rm(\n@@)}}
\def\deqno#1(#2){\eqno(\deq#1(#2))}
\def\deqal#1(#2){(\deq#1(#2))}
\def\eqback#1{{(\advance\eqNo by -#1 \lasteq)}}

\def\eqgroup(#1){{\global\advance\eqNo by1
       \edef\n@@{\lasteq}\global\atedef{E@#1}{\n@@}}}

\outer\def\iproclaim#1/#2/#3. {\vskip0pt plus50pt \par\noindent
     {\bf\dpcl#1/#2/ #3.\ }\begingroup \interlinepenalty=250\lessblank\sl}
\newcount\pcNo  \pcNo=0
\def\lastpc{\number\pcNo} 

\def\dpcl#1/#2/{\ifempty{#1}\global\advance\pcNo by1
       \edef\n@@{\lastpc}\else\edef\n@@{#1}\fi
       \ifempty{#2}\else\global\atedef{P@#2}{\n@@}\fi\n@@}
\def\pcl#1/#2/{\edef\n@@{#1}%
       \ifempty{#2}\else
       \ifatundef{P@#2}{\global\atedef{P@#2}{#1}}%
                       {\edef\n@@{\atname{P@#2}}}\fi
       \n@@}
\def\Def#1/#2/{Definition~\pcl#1/#2/}
\def\Thm#1/#2/{Theorem~\pcl#1/#2/}
\def\Lem#1/#2/{Lemma~\pcl#1/#2/}
\def\Prp#1/#2/{Proposition~\pcl#1/#2/}
\def\Cor#1/#2/{Corollary~\pcl#1/#2/}
\def\Exa#1/#2/{Example~\pcl#1/#2/}

\font\sectfont=cmbx10 scaled \magstep2
\def\secNo{00}
\def\Beginsection#1#2{\vskip\z@ plus#1\penalty-250
  \vskip\z@ plus-#1\bigskip\vskip\parskip
  \leftline{\bf#2}\nobreak\smallskip\noindent}
\def\bgsection#1. #2 .#3 \par{\Beginsection{.3\vsize}{\sectfont#1.\ #2 }%
            \def\secNo{#3}
            \eqNo=0
            }
\def\bgssection#1. #2 .#3 \par{\Beginsection{.3\vsize}{#1.\ #2 }%
            \def\secNo{#3}
            }
\def\bgsections#1. #2 \bgssection#3. #4 .#5\par{%
    \def\secNo{#5}
    \eqNo=0
        \Beginsection{.3\vsize}{\sectfont#1.\ #2 }
       \nobreak\leftline{\bf#3. #4}\par\noindent}
\def\Acknow#1\par{\ifx\REF\doref
     \Beginsection{.3\vsize}{\sectfont Acknowledgements}%
#1\par
     \Beginsection{.3\vsize}{\sectfont References}\fi}
\catcode`@=12

\overfullrule=0pt

\REF Fey Feyn \par
\REF HO HO \par
\REF Wo1 Woo \par
\REF Wo2 Woo2 \par
\REF Emc Emch \par
\REF GL GL \par
\REF GL2 GL2 \par
\REF BHW BHW \par
\REF BB+ hada \par
\REF MWS code \par
\REF Wig Wig \par
\REF Wig2 Wig2 \par
\REF Wig3 Wig3 \par
\REF JW JW \par

\line{} \vskip 2.0cm
\font\BF=cmbx10 scaled \magstep 3

{\BF \baselineskip=25pt \centerline{ Why Two Qubits Are Special} }
\vskip 1.0cm \ifx\draft1\centerline{  Version of \today }\fi
\vskip1cm

\centerline{
\bf
K.G.H. Vollbrecht and R.F. Werner%
}

\centerline{ {\sl Institut f\"ur Mathematische Physik, TU
Braunschweig,}}

\centerline{{\sl
Mendelssohnstr.3, 38106 Braunschweig, Germany}}%

\centerline{
{ \sl Electronic mail:\quad
   \sl k.vollbrecht@tu-bs.de , \sl r.werner@tu-bs.de  }}

\vskip 1.0cm

{ \narrower\narrower\noindent {\bf Abstract.}\
We analyze some special properties of a system of two qubits, and
in particular of the so-called Bell basis for this system, which
have played an important role in recent papers on entanglement of
qubits. In particular, we show which of these properties may be
generalized to higher dimension. We give a general construction
for bases of maximally entangled vectors in any dimension, but
show that none of the properties related to complex conjugation in
Bell basis can be realized for higher dimensional analogs. \par}
{\baselineskip=12pt \narrower\narrower\noindent \
 Pacs Nr: 03.67.-a , 03.65.-w , 03.65.Bz \

 \par}

\vskip20 pt

\vfil\eject

\bgsection I. Introduction .1

The qubit system is the smallest non-trivial quantum system.
Formerly known as a two-level system, it has often served as an
example for basic quantum phenomena\cite{Feyn}. Many of the basic
ideas of quantum information theory were first tested on qubits.
Indeed, for the invention of processes like entanglement enhanced
teleportation and dense coding it was very helpful to have an
explicit example, in which every detail could be explicitly worked
out. For these two processes the generalization to higher
dimensional systems was not difficult, hence the intuition gained
from the qubit case turned out to be valid.

On the other hand, in the theory of entanglement there have been
two achievements, which were so far only possible for qubits, and
probably have no higher dimensional analogs. These are the
``partial transpose'' \cite{HO} form of the criterion for separable
(classically correlated) states, and the remarkable formula of
Wootters\cite{Woo,Woo2} for the entanglement of formation of an
arbitrary state of two qubits. Hence in these cases it may be
dangerous to rely too much on the intuitions gained from the qubit
case. The purpose of this paper is to state as clearly as
possible, which of the ingredients of Wootters' proof and, in
particular, which properties of the ``Bell basis'' of
$\Cx^2\otimes\Cx^2$ have a chance of generalization to higher
dimensions.

Unfortunately, some crucial properties turn out to be specific to
two dimensions. We will show this by taking properties of the Bell
basis, stated in a form not referring to dimension, and proving
that the corresponding property can only hold in dimension two. We
hope that this will serve as a caveat and will help researchers in
the field to develop more accurate intuitions for higher
dimensional entanglement. As a first step we consider in Section~3
a property of the Bell basis which does generalize: it is an
orthonormal basis consisting of maximally entangled vectors. In
fact, up to a choice of phases and a local unitary transformation
the Bell basis is {\it uniquely} characterized by this property.
After collecting definitions and some basic properties of maximally
entangled vectors, we show this uniqueness, and give a fairly
general construction for bases of maximally entangled vectors,
which works in any dimension.

In Section~4 we look at Bell bases. Their most surprising and at
the same time most useful properties are two characterizations of
objects which have real components in this basis: real unitary
operators with determinant $1$ factorize, and real unit vectors
are maximally entangled. Our main result is that even weak forms
of either of these properties cannot be generalized to higher
dimensions.

\bgsection II. How single qubits are special .2

In this section we will describe some of the properties of single
qubits, which are false for systems with more than two dimensional
Hilbert space. Some of these are well-known, and we only recall
them because they are referred to later on. Others will have to be
treated in more detail for application in Section 4. Throughout we
will denote by $d$ the dimension of the Hilbert space of the
systems under consideration, so that qubits are characterized by
$d=2$.

Of course, everybody working in quantum information theory or
indeed quantum physics as a whole is familiar with the Poincar\'e
ball (or Bloch or Stokes sphere) representing the state space
(space of density matrices) of a two-level system. It is so well
circulated as the paradigm of a quantum state space that one must
perhaps warn students about its not so typical features. The most
conspicuous of these, which is in fact at the root of several
others, is that the ball has a {\bf center}. That is, there is a
density matrix $\Bar\rho=(1/2)\idty$ such that for every density
matrix $\rho$ there is an opposite one, $\rho'$, such that
$\Bar\rho=(\rho+\rho')/2$. In the language of Jordan algebras, an
axiomatic approach\cite{Emch} in which more exotic state spaces than usual
can arise, the $d\times d$-matrices are a ``spin factor'' iff and
only if $d=2$. Another geometrical feature
which is only valid in $d=2$ is that the extreme points (pure
states) form the complete (topological) {\bf boundary} of the
state space: in higher dimension every density matrix with {\it
some} zero eigenvalue is on the boundary, whereas the extreme
points are those with {\it all but one} eigenvalue equal to zero.

A consequence of the fact that for $d=2$ every one-dimensional
projection has only one one-dimensional projection in its
orthogonal complement is the failure of {\bf Gleason's Theorem}\cite{GL,GL2},
This Theorem says that for $d>2$ any real valued function on
one-dimensional projections, which sums to $1$ on every maximal
set of orthogonal one-dimensional projections, is given by the
expectations of a density matrix. Again, this has had some
repercussions in Axiomatic Quantum Mechanics.

Since in $d=2$ every pure state $\rho=\ketbra\phi\phi$ has
 a unique complement
it is natural to ask for a  ``Quantum
NOT'' operation\cite{BHW}, i.e., a map $\phi\mapsto\phi^\perp$,
which takes every vector $\phi\in\Cx^d$ to an orthogonal one,
$\phi^\perp$. It is easy to see that there can be no linear
operator $A$ such that $\phi^\perp=A\phi$: By definition, such an
operator would satisfy the equation $\bra\phi,A\phi>=0$ for all
$\phi$, from which one gets $A=0$ by polarization, i.e., by
inserting complex linear combinations for $\phi$. However, the
polarization trick uses complex linearity  in a crucial way, and
it is indeed possible to find {\it conjugate linear}
(``antilinear'') operators $\Theta$ such that
$$\bra\phi,\Theta\phi>=0 \quad,\deqno(Theta)$$ for all $\phi$.
Indeed, if $\Theta$ acts on the vectors of a basis
$\set{e_\alpha}$ as $\Theta
e_\alpha=\sum_\beta\Theta_{\beta\alpha}e_\beta$ then equation
\eq(Theta) is equivalent to
$\Theta_{\beta\alpha}=-\Theta_{\alpha\beta}$. Clearly, for $d>2$,
we have many choices for anti-symmetric matrices. A natural
additional requirement for a NOT operation would be that double
negatives should be the identity. It turns out that
$\Theta^2=\lambda\idty$ can hold for an anti-unitary NOT operation
only in even dimension (for odd $d$ an anti-symmetric matrix is
never invertible) and with $\lambda=-1$.

For $d=2$ there is only one antisymmetric matrix, up to a factor,
so the anti-unitary Quantum NOT operation is uniquely defined.
Because this argument works in any basis, we conclude that the
$\Theta$ is the same in every basis, so indeed this operation is
{\bf universal} in a very strong sense. Formally, this
universality is expressed by saying that
 $U\Theta U^*=\omega(U)\Theta$ for all ``basis changes'', i.e., all
unitaries $U$, where $\omega(U)\in\Cx$, $\abs{\omega(U)}=1$, is a
suitable phase. By looking at the universality condition in terms
of the matrix $\Theta_{\alpha\beta}$, one can see that for
$d\geq3$ a Universal NOT does not exist. However, the following
Proposition makes an even stronger claim: for $d\geq3$ there is no
universal anti-unitary at all.

\iproclaim/unot/Proposition. Let $d>1$ be natural number, and
suppose that there is a non-zero conjugate linear operator
$\Theta$ on $\Cx^d$ such that for any unitary operator $U$ there
is a phase $\omega(U)$ satisfying $U\Theta
U^*=\omega(U)\Theta$.\hfill\break Then $d=2$ and there is a factor
$\lambda\in\Cx$ such that $$ \Theta\pmatrix{a\cr
b}=\lambda\pmatrix{\Bar b\cr -\Bar a} \quad.\deqno(Theta2) $$
Moreover, $\omega(U)=\det(U)$, and  when $\abs\lambda=1$, $\Theta$
is anti-unitary. \eproclaim

\proof: Since $U\Theta U^*=\omega(U)\Theta$, we may take any
matrix element of this equation, which is non-zero for $\Theta$,
to conclude that $\omega(U)$ is a continuous function of the
matrix elements of $U$. Moreover, it is straightforward to verify
that $\omega$ is a character, i.e.,
$\omega(U_1U_2)=\omega(U_1)\omega(U_2)$.
Together with $\omega(\idty)=1$, this implies that
$\omega(U)=\det(U)^N$ for some integer $N$. Inserting
multiples of the identity, $U=\zeta\idty$ we find
$\zeta^2=\zeta^{Nd}$, i.e. $Nd=2$. Since we have assumed $d>1$,
this implies $N=1$ and $d=2$. That the $\Theta$ in \eq(Theta2)
satisfies the conditions  and is unique up to an factor was
already argued above.

A more elementary argument, not relying on the representation
theory of the unitary group, is the following.(We omit here the
part of the argument dealing with the null space of $\Theta$, so
assume $\Theta$ to be non-singular). Suppose that $u_1,\ldots,u_d$
are the eigenvalues of $U$ with eigenvectors $\phi_\alpha$, then
the vectors $\Theta\phi_\alpha$ are also eigenvectors with
eigenvalues $\omega(U)\Bar{u_\alpha}$.
Note that the conjugate $\Bar{ u_\alpha}$ appears, due to the
conjugate linearity of $\Theta$.
 It follows that the
spectrum of every unitary must be congruent to itself, subject to
a reflection followed by a rotation of the complex plane. For
$d=2$ the spectrum consists of two points on the unit circle, and
is hence symmetric with respect to a reflection on a line
orthogonal to $u_1-u_2$. Clearly, for $d\geq3$ a general set of
$d$ points on the unit circle has no such symmetry. \QED

Of course, the operator $\Theta$ from equation \eq(Theta2) also
satisfies the condition \eq(Theta).

\bgsection III. Maximally entangled states .3

We define a vector $\Phi\in\H_1\otimes\H_2$ in a Hilbert space
tensor product to be {\bf maximally entangled}, whenever both of
its restrictions are maximally mixed. The restricted density
matrices $\rho_1,\rho_2$ are defined by
$\tr(\rho_1A_1)=\langle\Phi,(A_1\otimes\idty)\Phi\rangle$ and
$\tr(\rho_2A_2)=\langle\Phi,(\idty\otimes A_2)\Phi\rangle$ for all $A_1,A_2$.Thus $\Phi$
is maximal entangled, if $\rho_1$ and $\rho_2$ are
proportional to the identities on $\H_1$ and $\H_2$.
By the well-known Schmidt decomposition this is only possible if
$\dim\H_1=\dim\H_2$, so we will set $\H_1=\H_2=\Cx^d$ throughout
this section. We will denote by $\maxent$ the set of maximally
entangled vectors.

The Schmidt decomposition for an arbitrary maximally entangled
vector $\Omega$ now reads
$$ \Omega={1\over\sqrt d}\sum_{\alpha=1}^d e_\alpha\otimes e'_\alpha
\quad,\deqno(Omega)
$$ where $\set{e_\alpha}$ and $\set{e'_\alpha}$ are suitable
orthonormal bases in the tensor factors. Generically (i.e., when
the reduced density matrices have only non-degenerate eigenvalues)
the Schmidt decomposition is unique up to phase factors. Maximally
entangled states constitute the opposite case: Since the reduced
density matrices are totally degenerate, the bases are only
determined up to a common unitary transformation, i.e.,
$$ \Omega=(U\otimes\Bar U)\Omega
\quad,\deqno(moveU)
$$
where $\Bar U$ is defined by the matrix elements $\langle
e'_\alpha,\Bar U e'_\beta\rangle
  =\Bar{\langle e_\alpha,Ue_\beta\rangle}$.
Note that this operation depends on the bases
$\set{e_\alpha},\set{e'_\alpha}$, and hence on the particular
maximally entangled state $\Omega$.

It is clear from the definition of maximal entanglement that local
unitary transformations, i.e., $U_1\otimes U_2$ with $U_1,U_2$
unitary, take maximally entangled vectors into maximally entangled
ones. In view of equation \eq(moveU) we get the same
transformations, by performing a unitary rotation only of one
factor. Other vectors, too, can be represented in this form. The
salient facts are collected in the following Lemma, the proof of
which relies completely on writing out everything in components
with respect to the bases $\set{e_\alpha}$ and $\set{e'_\alpha}$
appearing in \eq(Omega), and is left to the reader.

\iproclaim/XPhi/Lemma. %
Let $\Omega\in\Cx^d\otimes\Cx^d$ be a maximally
entangled unit vector. Then every vector
$\Phi\in\Cx^d\otimes\Cx^d$ can be written as
$$ \Phi=(X_\Phi\otimes\idty)\Omega
\quad,
$$
with a uniquely determined linear operator $X_\Phi$. Moreover,
\item{(1)} $\langle \Phi,\Psi\rangle=\tr(X_\Phi^*X_\Psi)$.
\item{(2)} $(X_\Phi\otimes\idty)\Omega =(\idty \otimes X^T_\Phi)\Omega$
\item{(3)} the restrictions of the state $\Phi$ are given by the density matrices
  ${1 \over d} X_\Phi^*X_\Phi$ and
  ${1\over d} \Bar X_\Phi X_\Phi^T$.
\item{(4)} $\Phi$ is maximally entangled iff $X_\Phi$ is
unitary.

$\Bar X_\Phi $ and $X^T_\Phi$ are defined by there matrix elements in the
basis given by the Schmidt decomposition of $\Omega$ \eq(Omega),
in the same way as the above complex conjugation \eq(moveU).
\eproclaim
How many maximally entangled states are there? Since the maximally
entangled vectors $\Phi$ are in on-to-one correspondence with the
unitaries $X_\Phi$, the manifold $\maxent$ has the same dimension
as the unitary group, i.e., $d^2$. But this says very little about
how $\maxent$ is embedded into $\Cx^d\otimes\Cx^d$. For example,
can we find an orthonormal basis of maximally entangled vectors?
In dimension $d=2$ the well-known Bell basis is an example: it
consists of the vectors
$$ \matrix{ \Phi_0=\hfill\Omega={1\over 2}(\ket\up\up>+\ket\down\down>)  &%
            \Phi_1=(i\sigma_1\otimes\idty)\Omega={i\over 2}(\ket\down\up>+\ket\up\down>) \cr%
            \Phi_2=(i\sigma_2\otimes\idty)\Omega={1\over 2}(\ket\down\up>-\ket\up\down>) &%
            \Phi_3=(i\sigma_3\otimes\idty)\Omega={i \over 2}(\ket\up\up>-\ket\down\down>)}
\quad.\deqno(Bbasis)$$
The factors $i$ are, of course, irrelevant at this stage, but will
turn out to be crucial for the further properties of this basis
studied in the next section.

\bgssection III.2. Constructing Bases of maximally entangled vectors .3

By Lemma 1 the task of constructing a basis
$\set{\Phi_\alpha}\subset\maxent$ is equivalent to finding a basis
of unitary operators $X_\alpha$ on $\Cx^d$, satisfying the
orthonormality condition
$$ \tr(X_\alpha^*X_\beta)=d\, \delta_{\alpha\beta}
\quad,\quad \alpha,\beta=1,\ldots,d^2.\deqno(Ubasis)
$$
It turns out that such bases exist in any dimension. Indeed, a
rough dimension count indicates that there should be many bases of
this kind: the manifold of $d^2$-tuples of maximally entangled
vectors has dimension $d^2(d^2-1)$, where we subtracted $1$ for
the phase ambiguity in each basis vector. The only remaining
conditions are the $d^2(d^2-1)/2$ orthogonality conditions between
different vectors. If we want to identify bases which can be
transformed into each other by local unitaries we should subtract
furthermore the  dimension of this group, $2d^2-1$. So we are left
with  a dimension count for the manifold of maximally entangled
bases growing in leading order like $d^4/2$.

There are several general constructions for bases of unitaries.
Since such bases are precisely what is needed for generalization
of the entanglement enhanced teleportation scheme to dimensions
$d>2$, one such construction (working for any $d$) has been noted
in \cite{hada}. Here we give the most general construction known to
us. A {\bf Hadamard matrix} $H$ is, by definition, a square
matrix, in which all entries have modulus one, and which is
unitary up to a factor: $$ \abs{H_{k\ell}}=1\quad,\quad
k,\ell=1,\ldots,d \quad,\quad HH^*=d \idty\quad.\deqno(Hadamard)
$$

From this we construct what we call {\bf shift-and-multiply bases}
of $d^2$ unitary operators $\set{U^{ij}}_{i,j=1\dots d}$. The
construction depends on a collection of $d$ Hadamard matrices
$H^j$ of dimension $d\times d$, and a {\it bi-injective
composition} $\tau:I_d\times I_d\to I_d$, where
$I_d=\set{1,\ldots,d}$. This composition need be neither
commutative nor associative, but we require ``bi-injectivity'',
defined as the cancellation laws $(\tau(i,k)=\tau(j,k)
)\Rightarrow(i=j)$ and $(\tau(k,i)=\tau(k,j) )\Rightarrow(i=j)$.
In other words, every symbol appears exactly once in each row or
column of the composition table. With $\set{e_k}_{k=1}^d$ the
canonical basis of $\Cx^d$, we define the operators
$$  U^{ij}e_k= H_{ik}^je_{\tau(k,j)}
\quad.\deqno(Hbasis)
$$
We leave to the reader the verification that in this way any
collection of Hadamard matrices $H^j$ and a composition $\tau$
generates a orthonormal basis of unitaries, and hence a basis of maximally
entangled vectors.

The problem is now shifted to constructing Hadamard matrices. For
the case of real entries ($H_{k\ell}=\pm1$) this is a well-known
problem arising in coding theory\cite{code}. Several families are
known, but no general construction. The complex case is less
well-studied. A simple construction is based on the theory of
(finite) abelian groups: If $G$ is an abelian group of order $d$,
then there are exactly $d$ different characters, i.e., mappings
$\gamma:G\to\Cx$ such that $\abs{\gamma(g)}=1$, and
$\gamma(g_1g_2)=\gamma(g_1)\gamma(g_2)$. The set of characters is
called the dual group $\Gamma$, and the Fourier transform ${\cal
F}$ takes functions on $G$ into functions on $\Gamma$ via $({\cal
F}f)(\gamma)=\sum_g \gamma(g)f(g)$. It is well known that this
transformation is unitary up to a normalization factor. Hence the
$d\times d$ coefficients $\gamma(g)$ for $\gamma\in\Gamma, g\in G$
form a Hadamard matrix. The simplest choice of this kind (which
works in any dimension $d$) is based on the cyclic group of order
$d$, which is its own dual. The associated Fourier matrix is then
$$ H_{k\ell}=\exp\left({2\pi i\over d}k\ell\right)
\quad.\deqno(Hcycl) $$ The direct product of groups in this
construction leads to the tensor product of Hadamard matrices.
More generally, the tensor product for Hadamard matrices is again
a Hadamard matrix. Similarly, the tensor product of bases of
unitaries (or maximally entangled states) is again a basis of the
required type. Thus whenever $d$ is a composite number, bases can
be constructed from bases of smaller dimensions (the factors of
$d$). The only abelian groups of prime order are the cyclic
groups. For $d=2,3$ this also leads to the only Hadamard matrix (up
to trivial transformations). However, already for order $5$ there
are Hadamard matrices not arising from the cyclic group, so the
shift-and-multiply construction of bases of unitaries is strictly
more general than the one based on abelian groups. On the other
hand, the dimension count described above suggests, that the
shift-and-multiply construction is still not the most general one.
In fact, it seems to be an open problem to characterize all bases
of unitaries for $d=3$. Only for $d=2$ the Bell basis is
essentially the only basis of maximally entangled vectors.

\iproclaim/Bellunique/Lemma. %
Let $\{\Psi_\alpha\}, \alpha=0 \dots 3$ be a maximal entangled
basis of  $\Cx^2\otimes \Cx^2$ and let $X_\alpha$ denote the
unitaries such that $\Psi_\alpha=(X_\alpha\otimes\idty)\Omega.$
\item{(1)} Then there are unitaries $U_1,U_2$ and phases $\phase_\alpha$
such that $\Psi_\alpha=\phase_\alpha (U_1 \otimes U_2) \Phi_\alpha$,
where $\{\Phi_\alpha\}$ denotes the standard Bell basis
\eq(Bbasis).
\item{(2)} If all $X_\alpha$ have the same determinant then either
 (a) the phases $\phase_\alpha $ may all be chosen equal to $1$ or
 (b) the phases may be chosen as $\set{1,1,-1,1}$ or, equivalently,
     $U_1 \otimes U_2 \Phi_\alpha$ may be made into an odd permutation
      of the given basis.
\eproclaim

\proof: From Lemma $2.(2)$ a local unitary transformation
of the vector $\Psi_\alpha$ affect the $X_\alpha$ matrix like
$$U_1 \otimes U_2 \Psi_\alpha
   =((U_1 X_\alpha U_2^T)\otimes \idty)\Omega
\quad,   \deqno()$$
 i.e., $X_\alpha\mapsto (U_1 X_\alpha U_2^T)$.
By choosing  $U_1=\idty$ and $U_2=X_0^*$, we may assume that
$X_0=\idty$. Note that under the assumptions of part (2) this also
achieves $\det X_\alpha=1$ for all $\alpha$. The local unitary
transformations leaving the condition $X_0=\idty$ invariant are
 $X_\alpha\mapsto UX_\alpha U^*$. Moreover, from orthogonality
with $X_0$ we get $\tr(X_\alpha)=0$. This means that each of the
unitaries $X_\alpha$, $\alpha=1,2,3$ has two eigenvalues adding up
to $0$, and is hence of the form
 $X_\alpha=i\phase_\alpha \vec r_\alpha\cdot\vec\sigma$, where the
$\vec r_\alpha\in\Rl^3$ are real three dimensional unit vectors,
and $\vec\sigma$ is the vector of Pauli matrices. Orthogonality of
the $\Psi_\alpha$ implies these three vectors to be orthogonal,
too. Moreover, condition (2) is equivalent to
$\phase_\alpha=\pm1$, or $\phase_\alpha=1$, since a sign can be
absorbed in $\vec r_\alpha$.

Since the operation $X\mapsto UXU^*$ is just a three dimensional
proper rotation, we can rotate the orthonormal frame $(\vec
r_1,\vec r_2,\vec r_3)$ to be parallel to the standard basis in
$\Rl^3$. Hence we get $X_\alpha=i\phase_\alpha\sigma_\alpha$,
proving part (1), or $X_\alpha=\pm i\sigma_\alpha$ in case (2). By
further rotations we can make all signs but at most one $+1$. The
distinction between cases (a) and (b) is precisely, whether the
real orthogonal transformation taking the frame $(\vec r_1,\vec
r_2,\vec r_3)$ to the standard basis has determinant $+1$ or $-1$.
In the second case we need an orientation reversing operation
(such as reversing one direction or permuting some basis elements)
before a proper rotation (implemented by a local unitary) brings
the given basis to the standard form. \QED

\bgssection III.3. Unitaries respecting maximal entanglement .3

As noted above, all local unitaries map the set $\maxent$ of
maximally entangled vectors into itself. It turns out that the
converse is also true, apart from one obvious counterexample, the
``flip'' unitary defined by
$\Flip(\phi\otimes\psi)=\psi\otimes\phi$:

\iproclaim/UM=M/Proposition. %
Let $U$ be a unitary operator on $\Cx^d\otimes \Cx^d$. Then
$U\maxent\subset\maxent$ if and only if $U$ is local up to a flip,
i.e., there are unitaries $U_1,U_2$ such that either
 $U=U_1\otimes U_2$ or $U=(U_1\otimes U_2)\Flip$.
\eproclaim

\proof:
 Every unitary operator  $U$ on $\Cx^d\otimes \Cx^d$ defines a linear
bijective map $f_U$ from the space of all $d\times d$-matrices
into itself by
$$U \Phi= U(X_\Phi\otimes\idty)\Omega
   =: (f_U(X_\Phi)\otimes\idty)\Omega .
\deqno(defi)$$
 Then by \Lem/XPhi/\ the condition $U\maxent\subset\maxent$ is
equivalent to $f_U$ taking unitary operators to unitary
operators. We have to show that in this case $f_U$ can be written
either as $f_U(X)=U_1 X U_2$ or $f_U(X)=U_1 X^T U_2$ , which is
equivalent to above proposition. From \Lem/XPhi/ it is easy to see
that in this sense the transposition belongs to the flip
operation: $f_\Flip(X)=X^T$. We note that since the implication $U\maxent\supset\maxent$ is
trivial, the assumption $U\maxent\subset\maxent$ is actually
equivalent to $U\maxent=\maxent$, and hence also to
$U^*\maxent\subset\maxent$.

By composing $U$ with a local unitary map we may assume that
$f_U(\idty)=\idty$ or, equivalently that the reference  vector $\Omega$
from Lemma $2$
is invariant under $U$. Consider a
unitary $X=e^{iA}=\idty + i A - {1 \over 2} A^2 + O(A^3)$ close to
the identity (with $A=A^*$ small). Then $f_U(X)$ also has to be
unitary, and we will evaluate this condition to second order in
$A$, using the linearity of $f_U$: $$\eqalign{
  \idty &=  f_U(X)^* f_U(X) \cr
        &= \idty+f_U(i A)^* + f_U(i A) -
           {1 \over 2}(f_U(A^2)+f_U(A^2)^*)+f_U(i A)^*f_U(iA)+O(A^3).
}\deqno(gammel2)$$
 From the first order, $f_U(A)=f_U(A)^*$ for $A=A^*$. Hence from the second order
$f_U(A^2)={1 \over 2}(f_U(A^2)+f_U(A^2)^*)=f_U(i A)^*f_U(iA)$ is a
positive operator. Since every positive operator can be written as
$A^2$ for some $A=A^*$, we find that $f_U$, and by the same token
its inverse, map positive operators to positive operators. Hence by
Wigner's Theorem \cite{Wig} $f_U$ can be written either as $f_U(X)=S X S^*$
or $f_U(X)=S X^T S^*$ with some unitary $S$.
(We use a form of Wigner's Theorem with is formulated with positiv and invertible
maps. \cite{Wig2,Wig3} )
 \QED

\bgsections IV. Bell bases
\bgssection IV.1. Characterization Theorem .4

The most surprising properties of the Bell basis are related to
the anti-unitary operation of complex conjugation in this basis: a
vector is maximally entangled iff its components with respect to
the Bell basis are real up to a factor, and a unitary operator on
$\Cx^2\otimes\Cx^2$ is local iff, after multiplication with a
suitable phase, it becomes real in Bell basis and has determinant
$+1$ ({In the folklore on this subject, the
determinant condition is sometimes forgotten, inviting the flip
$\Flip$ as an obvious counterexample, see \Prp7/Udet/}). Both
these properties are extremely useful and play a crucial role in
the Wootters-formula for the two qubit system. It is thus highly
desirable to find extensions to higher dimensional systems. One
direction in which a generalization might be sought is to break
the above ``iff'' statements, and to require in higher dimensions
maybe only one direction of implication. This leaves four
possibilities to be tested. However, as the following Theorem
shows, none of them can be realized in any dimension $d>2$.

\iproclaim/theorem/Theorem. %
Let $d\in\Nl$, and $\Psi_\alpha, \alpha=0,\ldots,d^2-1$ a basis of
maximally entangled vectors in $\Cx^d\otimes\Cx^d$. Let $X_\alpha$
denote the unitaries such that
$\Psi_\alpha=(X_\alpha\otimes\idty)\Omega$, with a maximal
 entangled vector $\Omega$ (see \Lem/XPhi/). Then
the following conditions are equivalent:
 \item{(1)} $d=2$, and there are unitary operators $U_1,U_2$ and a
    permutation $\pi$ such that the $\Psi_\alpha$ can be written as
 $(U_1\otimes U_2)\Psi_\alpha=\Phi_{\pi(\alpha)}$
 with $\alpha=0,\ldots,3$.
 \item{(2)} $U_1,U_2 \in SU_d$ $\Rightarrow
      \forall_{\alpha,\beta}\langle \Psi_\alpha ,
                U_1 \otimes U_2 \Psi_\beta\rangle \in \Rl $
 \item{(3)} $U \in SU_{d^2}$ and $\forall_{\alpha,\beta}
       \langle \Psi_\alpha , U\Psi_\beta\rangle \in \Rl$
       $\Rightarrow\quad U=U_1 \otimes U_2$ or $U=(U_1 \otimes U_2) \Flip$
 \item{(4)} $\phi \in \maxent$  $\Rightarrow \exists_{\omega, |\omega|=1}\forall_{\alpha}
            \omega \langle \Psi_\alpha,\phi\rangle \in \Rl$
 \item{(5)}$\forall_{\alpha} \langle \Psi_\alpha,\phi\rangle \in \Rl$ $\Rightarrow \phi \in \maxent$

 \item{(6)} $X_\alpha^*X_\beta+X_\beta^*X_\alpha=2\,\delta_{\alpha\beta}\idty$.
\eproclaim

 \proof:
We will prove the inclusions:
 $(1)\Rightarrow(2)\Rightarrow(4)\Rightarrow(5)\Rightarrow(6)\Rightarrow(1)$, and
$(4 \hbox{\ and\ }5)\Rightarrow(3)\Rightarrow(5)$.

$(1)\Rightarrow(2)$ It suffices to take $\Psi_\alpha=\Phi_\alpha$
as the standard Bell basis \eq(Bbasis) with $\Phi_0=:\Omega$
and $\Phi_\alpha=(i\sigma_\alpha\otimes\idty)\Omega$,
($\alpha=1,2,3$). Since exponentiation is a power series with real
coefficients, it suffices to show that the generators of the local
unitary group with determinant one, namely $i\sigma_k\otimes
\idty$ and $\idty \otimes i\sigma_k$ ($k=1,2,3$) are real in the
standard Bell basis. Computing the matrix elements
 $\langle\Phi_\alpha, (i\sigma_k\otimes\idty)\Phi_\beta\rangle$
of the generators involves a case distinction as to how many of
$\alpha,\beta$ are equal to $0$. If $\alpha=\beta=0$ we get
 $\langle\Phi_0,  (i\sigma_k\otimes\idty)\Phi_0\rangle
 =d^{-1}\tr(i\sigma_k)=0$,
because $\Phi_0$ is maximally entangled, and its restriction to
the first factor is ${1\over d}\idty$. If exactly one of
$\alpha,\beta$ is zero, the matrix element carries an even power
of $i$, and we get matrix elements of the form
 $\langle\Phi_0,  (i\sigma_k\otimes\idty)\Phi_l\rangle
 =-d^{-1}\tr(\sigma_k\sigma_l)$, which is real anyway.

 If both are non-zero, we find
$$\langle \Phi_\alpha , i \sigma_k\otimes \idty  \quad
\Phi_\beta \rangle =\langle \Omega,i(\sigma_\alpha \sigma_k
\sigma_\beta)\otimes \idty \quad  \Omega \rangle ={i \over 2}
tr(\sigma_\alpha \sigma_k \sigma_\beta). \deqno()$$
 where $\alpha,\beta,k=1 \dots 3$. When two indices are the same
this trace is zero, when they are all different, the relations
$\sigma_1\sigma_2=i\sigma_3$ (and cyclic) imply that the trace is
imaginary and the matrix element is real.

$(2)\Rightarrow (4)$
From \eq(Omega) it is easy to see, that
all maximally entangled vectors are
 equivalent by local unitary transformation. So every maximally entangled vector
$\phi$ can be written as $\phi=\Bar\omega  (U_1 \otimes U_2)
\Psi_0$, with  $U_1,U_2 \in  SU_d$ and a phase $\Bar\omega$.
$\omega \langle \Psi_\alpha,\phi\rangle$ is hence
 the $(\alpha,0)$-row of
matrix elements of $U_1 \otimes U_2$  in the $\{ \Psi_\alpha \}$
basis.
Condition $(2)$ guarantees that these matrix elements are
real.

$(4) \Rightarrow (5)$ Condition (4) refers to two different sets
of vectors in $\Cx^d\otimes\Cx^d$: on the one hand, the space of
maximally entangled vectors $\maxent$, which by \Lem/XPhi/ can be
parameterized by the unitary group $U_d$, and on the other hand
the space of ``up to an overall phase factor real in $\{\Psi_\alpha\}$-basis'' normalized
vectors, which we call ${\cal Q}$ for the sake of this proof. So
(4) means $\maxent\subset {\cal Q}$ and we now have to show ${\cal Q} \subset\maxent$ .
These two manifolds of
vectors have the same dimension, namely $d^2$: On the one hand
this is the dimension of $U_d$ (the tangent space at the identity
is the space of hermitian operators). On the other hand, a real
vector has $d^2$ real components. The overall phase for vectors in
${\cal Q}$ adds an extra dimension, but we have to subtract one
for normalization.

Now consider a small neighborhood $N\subset\maxent$ of some point $\Phi\in\maxent$.
We can parametrize its points uniquely as $(U\otimes\idty)\Phi$,
with $U$ in a neighborhood of the identity in $U_d$. Thereby we
get a $d^2$-dimensional set of vectors, which by assumption (4) lies
in the $d^2$-dimensional  manifold ${\cal Q}$, and hence contains
an open neighborhood of $\Phi$ in ${\cal Q}$.  This shows that
$\maxent$ is an open subset of ${\cal Q}$. On the other hand,
$\maxent$ is the continuous image of the compact space $U_d$,
hence compact, hence closed in ${\cal Q}$. But ${\cal Q}$ is
clearly connected. So $\maxent$, being both open and closed, must
be equal to ${\cal Q}$.

$(5) \Rightarrow (6)$ Condition (5) means that every vector of the
form
$$\phi=\sum_\alpha a_\alpha \Psi_\alpha
   =( (\sum_\alpha a_\alpha X_\alpha) \otimes\idty) \Omega
\deqno()$$
 with real $a_\alpha, \sum_\alpha a_\alpha ^2=1$ is maximally entangled.
Therefore $\sum_\alpha a_\alpha X_\alpha$ has to be unitary for
every normalized real vector $\vec{a}$. Expanding the unitarity
condition, and using the normalization condition to cancel the
diagonal, we are left with the condition
$$\sum_{\alpha > \beta} a_\alpha a_\beta \quad
     (X_\alpha^*X_\beta+X_\beta^*X_\alpha) =0
\quad .\deqno()$$
 Since this  holds for all vectors $\vec{a}$ each term
of this sum has to be zero. The relation for $\alpha=\beta$ is
clear from the unitarity of each $X_\alpha$.

$(6)\Rightarrow(1)$ Note that unitaries satisfying these relations
retain this property under the transformation $X_\alpha\mapsto
UX_\alpha$, with $U$ unitary. Choosing $U=X_0^*$, we find that we
may assume $X_0=\idty$ without loss of generality. Then the
relations for $\beta=0$ say that $X_\alpha+X_\alpha^*=0$. Setting
$X_\alpha=iR_\alpha \quad\alpha>0$, the problem is reformulated to finding
$d^2-1$ hermitian, unitary, operators acting on a $d$-dimensional
Hilbert space satisfying the relations \eq4.4(cliff) below. Hence
by \Lem6/cliffLem/, $d$ is even, $N=d^2-1$ is odd, and hence
$d=2^{d^2/2-1}$. This is possible only for $d=2$. We can thus
invoke \Lem/Bellunique/ showing that the $R_\alpha$ must be the
Pauli matrices, up to at most a permutation of the indices.

$(4 \hbox{\ and\ }5)\Rightarrow(3)$
 From (4) and (5) it follows, that a unitary matrix, which is
real in the $\set{\Psi_\alpha}$ basis maps $\maxent$ into itself.
Hence (3) follows from \Prp/UM=M/.

$(3)\Rightarrow (5)$
 Any unit vector $\phi$ which is real in some basis $\set{\Psi_\alpha}$
can be obtained by rotating the first basis vector $\Psi_0$ in
his direction via a in this basis real orthogonal transformation. This is to
say that there is a unitary operator $U$ satisfying the hypothesis
of (3) and $\phi=U\Psi_0$. Hence, whether $\phi=(U_1\otimes
U_2)\Psi_0$ or  $\phi=\Flip (U_1 \otimes U_2)\Psi_0$, this vector
is maximally entangled.\QED

To complete the proof, especially the crucial step
$(6)\Rightarrow(1)$, in which dimension $d=2$ is forced, we
invoked the following Lemma, which belongs to the representation
theory of {\bf Clifford algebras}. It can be found, e.g., in
\cite{JW}. But since it is a crucial step, we will give an
independent proof in the following Lemma.

\iproclaim/cliffLem/Lemma. Assume that $R_1,\ldots,R_N$ is a set
of $N>1$ hermitian operators(generators) acting irreducibly on a
$d$-dimensional space, and satisfying the relations
$$R_\alpha R_\beta + R_\beta R_\alpha  =2\delta_{\alpha\beta}\idty
.\deqno(cliff)$$
 Then $d$ is even, and if $N$ is odd, we have $d=2^{(N-1)/2}$.
\eproclaim
\proof:
Cause this Lemma belongs to the representations theories of algebraic
groups, we will now denote $R_\alpha$ as the generators of a
group.
Consider the generator $R_1$: Setting $\alpha=\beta=1$ in \eq(cliff)
it can be seen, that $R_1$ has two eigenspaces for
the eigenvalues $\pm1$, and from the relation
 $R_\alpha R_1=-R_1R_\alpha$ it is clear that each of the other generators
exchanges these two eigenspaces. Since $R_\alpha$ is unitary, this
also shows that the eigenspaces are of equal dimension, so $d$ is
even. Let us take the second generator, $R_2$ to furnish a
standard mapping between these spaces. Then we can characterize
the action of generators $R_\alpha$ with $\alpha\geq3$ completely
by the action of $R_2R_\alpha$ inside the ``$+1$''-eigenspace of
$R_1$. In other words, we consider, for $\alpha\geq3$ the
operators
$$ R_\alpha'= i R_2R_\alpha$$
It is straightforward to verify that these operators again satisfy
the Clifford relations \eq(cliff), and are Hermitian. Moreover, they commute with
$R_1$. Restricting to the ``$+1$''-eigenspace of $R_1$ we are thus
left with the same representation problem as before, albeit with
$N'=N-2$ generators, and in a representation space of dimension
$d'=d/2$. Moreover, the $\set{R_\alpha'}$ are again an irreducible
set, because any operator $C'$ commuting with them all determines
an operator $C$ commuting with the $R_\alpha$, by extending $C'$
as $R_2C'R_2$ to the ``$-1$''-eigenspace of $R_1$.

This argument can be iterated until exactly one generator is left
(since $N$ is odd). Irreducible representations of the only
remaining relations $R_N^2=\idty, R_N=R_N^*$ are one-dimensional,
with $R=\pm1$. (The sign coming out at this stage can also be
determined from the sign of the product $R_1R_2\cdots R_N$, which
commutes with all $R_\alpha$ by virtue of \eq(cliff), and is hence
a multiple of the identity). Collecting the factors $2$  for the
dimension then gives $d=2^{(N-1)/2}$. \QED

We now want to look back at condition $(3)$ of Theorem 5. It
would be nice here to have a simple condition on $U$
distinguishing the two cases. It turns out that this criterion is
simply the determinant of $U$.
\iproclaim/Udet/Proposition. %
Let $U$ be a unitary operator on $\Cx^2\otimes \Cx^2$ which is
real in the standard Bell basis. Then $U= (U_1 \otimes U_2)$ iff
$\det(U)=1$ and $U= (U_1 \otimes U_2) \Flip$ iff $\det(U)=-1$.
\eproclaim

\proof: From \Thm/theorem/ we know that $U$ has to factorize in
one of the given forms. $U$ lies in the connected component of the
identity of $SO_4$ iff its determinant is one, and iff it can be
written as the square of another element, say $U=V^2$. Either
factorization for $V$ now implies that $U$ is local. Since $SU_2$ is
connected, every local unitary $U$ can be written as a square, so
$\det U=1$. This proves the first assertion, and hence the
remaining cases, $\det U=-1$ and $U\Flip$ local must also match.
Indeed, $\det\Flip=-1$, because the dimension of its
``$-1$''-eigenspace is one, hence odd. \QED
\vskip 0pt plus1800pt

\bgssection IV.2. Conjugation in Bell basis .4

The remarkable properties of the Bell basis described in
\Thm/theorem/ are in some sense not so much a property of this
basis, but of the anti-unitary operation of {\it complex
conjugation} in Bell basis. Indeed, if we change the Bell basis by
a local unitary transformation, the new basis vectors will also be
maximally entangled, hence real in Bell basis (up to a common
factor), and the complex conjugation with respect to the new basis
will be exactly the same as before (again, up to a common factor).
Hence this conjugation operation is ``universal'' in a way very
similar to the Universal NOT of \Prp/unot/.
We would like to formulate the following Proposition in a more
general way, so that it also could be applied  to
multi-particle-systems:

\iproclaim/BellConj/Proposition. Let $d_1,d_2\dots d_n>1$,$n\geq2$ be natural
numbers, and suppose that there is a non-zero conjugate linear
operator $\Theta_n$ on $\Cx^{d_1}\otimes\Cx^{d_2}\dots \otimes \Cx^{d_n}$ such that for
any local unitary operator $U=U_1\otimes U_2 \dots \otimes U_n$ there is a phase
$\omega(U)$ satisfying
$U\Theta_n U^*=\omega(U)\Theta_n$.\hfill\break
 Then $d_1=d_2 \dots=d_n=2$ and there is a factor $\lambda\in\Cx$ such that
$$ \Theta_n=\lambda\Theta\otimes\Theta \dots \otimes\Theta=: \Theta^{\otimes n}$$
and $\omega(U)=\det(U)$, where $\Theta$ denotes the operator described in \Prp/unot/.
For  two qubits we get $\Theta_2= \lambda \Theta \otimes \Theta  = \lambda\Theta_{\rm
Bell}$, where  $\Theta_{\rm Bell}$ denotes the complex conjugation in Bell basis.
\eproclaim

Note that the antilinear operator-tensor-product is uniquely defined on
product vectors and from there it can be extended by antilinearity
to arbitrary vectors.

\proof:
Similarly to the proof of \Prp/unot/ we get that $\omega(U)$ is
a character, i.e., $\omega(U_1 U_2)=\omega(U_1) \omega(U_2)$.
Therefore, it is clear that $\omega(U)$ has to factorize in
the following form: $\omega(U_1\otimes U_2 \dots \otimes
U_n)=\omega_1(U_1)\omega_2(U_2)\dots\omega_n(U_n)$.
$d_1=2$ follows by applying exactly the same arguments as
in the proof of \Prp/unot/ to the equation
$(U_1\otimes\idty \dots \otimes \idty)\Theta_n(U_1\otimes\idty\dots \otimes\idty)^*=\omega_1(U_1)\Theta_n$.
Similarly, we get $d_{2}, \dots,d_n=2$ and $\omega(U)=\det(U)$. It is clear that
$\Theta_n=\Theta\otimes\Theta \dots \otimes \Theta$ has the required properties. On the
other hand, if $\Theta_n$ and $\Theta_n'$ both satisfy these
conditions, the {\it linear}   operator $C=\Theta_n\Theta_n'$
satisfies the equation $UCU^*=\omega(U)\overline{\omega'(U)}C=|\det(U)|^2
C=C$.
Therefore $C$ commutes with all local unitaries and
has to be a multiple of the identity, and therewith $\Theta^2_n$ and
${\Theta'}^2_n$ are multiple of the identity and finely $\Theta'_n$ is
a multiple of $\Theta_n$.

For the two qubit-system ($n=2$) this shows that
$\Theta_{\rm Bell}=\lambda\Theta\otimes\Theta$. However, the Proposition makes the
stronger claim that $\lambda=1$. This is
readily verified by checking that the Bell basis is invariant
under $\Theta\otimes\Theta$.
\QED

Squaring the equation in the proposition, and using
anti-unitarity, we find
$$\Theta_n^2=|\lambda|^2(\Theta^2)^{\otimes n}
   =|\lambda|^2\ (-1)^n\ \idty,
$$
where $|\lambda|=1$ characterizes the unitary case.
From this we see that the $\Theta_n$-operation applied to
density matrices ($\rho \mapsto \Theta_n
\rho\Theta_n^*$), as it is used in the Wootters formula for two qubits,
can have pure fixed points only if $n$ is even.
Exactly in these cases $\Theta_n$ can be
identified with the complex conjugation in some basis, namely
the tensor product of the Bell bases.
On the other hand, if $n$ is odd, $\Theta_n$ is a NOT-operation in
the sense of Section~2, although, of course, not a universal one with
respect to unitaries other than local ones. That is
$\bra\Psi,\Theta_n\Psi>=0$ for {\it all} vectors, not just for
product vectors. In either case, no application of $\Theta_n$ to multi-particle
entanglement is known to us.

\let\REF\doref
\Acknow KGHV acknowledges support from the Deutsche
Forschungsgemeinschaft, program QIV.

\REF Fey Feyn \Bref
    R.P. Feynman, R.B. Leighton, M. Sands
    "The Feynman lectures on physics" Vol.III, Addison-Wesley,
    Reading 1966

\REF HO HO \Jref
   M. Horodecki, P. Horodecki, H. Horodecki
    "Separability of mixed states: necessary and sufficient conditions"
    Phys.Lett.A. @233(1996)1--8

\REF Wo1 Woo \Jref
     W.K. Wootters
    "Entanglement of formation of an arbitrary state of two
    qubits"\hfill\break
    Phys.Rev.Letters @80(1998) 2245--2248

\REF Wo2 Woo2 \Jref
    S. Hill,W.K. Wootters
    "Entanglement of a pair of quantum bits"
    Phys.Rev.Lett. @78(1997) 5022--5025

\REF Emc Emch \Bref
    G.G. Emch
    "Algebraic methods in Statistical Mechanics and Quantum Field Theory"
    Wiley 1972

\REF GL GL \Jref
   A.M. Gleason
   "Measures on the closed subspaces of a Hilbert space"
    J.Math.Mech @6(1957)885--893

\REF GL2 GL2 \Jref
   R. Cooke, M. Keane, W. Moran
   "An elementary proof of Gleason's theorem"
    Math.Proc.Comb.Phil.Soc.  @98(1985)117

\REF BHW BHW \Gref
    V. Buzek, M. Hillery, R.F. Werner
    "Optimal manipulation with qubits: Universal NOT gate"
    Report quant-ph/9901053 , to appear in Phys.Rev. A

\REF BB+ hada \Jref
    C.H. Bennett, G. Brassard, C. Crepeau , R. Jozsa, A. Peres,
    W.K. Wootters
    "Teleporting an unknown quantum state via dual classical and Einstein-Podolsky-Rosen channels"
    Phys.Rev.Lett. @70(1993)1895--1899

\REF MWS code \Bref
    F.J. MacWilliams, N.J.A. Sloane
    "The theory of error-correcting codes I,II"
    North-Holland, Amsterdam 1977

\REF Wig Wig \Bref
    E.P. Wigner
    "Gruppentheorie und ihre Anwendung auf Quantenmechanik der
    Atomspektren"
    Vieweg 1931

\REF Wig2 Wig2 \Jref
    R.V. Kadison
    "Isometries of operator algebras"
    Annals of Math. @54(1951) 325--338

\REF Wig3 Wig3 \Bref
    E.P. Daves
    "Quantum theory of open systems"
    Academic Press 1976

\REF JW JW \Jref
   P.E.T. J{\o}rgensen, R.F. Werner
    "Coherent states of q-canonical commutation relations"
    Commun.Math.Phys. @164(1994)455--471

\end